\documentclass[preprint,onecolumn,superscriptaddress,12pt]{revtex4-1}

\usepackage{graphicx,natbib}
\usepackage{dcolumn}
\usepackage{bm}
\usepackage{amsmath,graphics,amssymb,psfrag}
\usepackage[colorlinks=true,linkcolor=blue,urlcolor=blue,citecolor=blue]{hyperref}
\usepackage[switch]{lineno}
\usepackage{tikz,xcolor,hyperref}
\definecolor{lime}{HTML}{A6CE39}
\DeclareRobustCommand{\orcidicon}{%
    \begin{tikzpicture}
    \draw[lime, fill=lime] (0,0)
    circle [radius=0.16]
    node[white] {{\fontfamily{qag}\selectfont \tiny ID}};\draw[white, fill=white] (-0.0625,0.095)
    circle [radius=0.007];
    \end{tikzpicture}
    \hspace{-2mm}}
\foreach \x in {A, ..., Z}
{\expandafter\xdef\csname orcid\x\endcsname{\noexpand\href{https://orcid.org/\csname orcidauthor\x\endcsname}{\noexpand\orcidicon}}}

\begin{document}

\title{Radiative recombination rate suppressed in a quantum photocell with three electron donors}

\author{Jing-Yi Chen }
\affiliation{Department of Physics, Faculty of Science, Kunming University of Science and Technology, Kunming, 650500, PR China}

\author{Shun-Cai Zhao\orcidA{}}
\email[Corresponding author: ]{zsczhao@hotmail.com }
\affiliation{Department of Physics, Faculty of Science, Kunming University of Science and Technology, Kunming, 650500, PR China}



\begin{abstract}\linenumbers
The radiative recombination of electron-hole pairs represents a great challenge to the photon-to-charge efficiency in photocell. In this paper, we visit the radiative recombination rate (RRR) in a quantum photocell with or without three dipole-dipole coupled electron donors. The results show that different gaps play the same roles while the ambient temperatures play different roles in the suppressed RRR with or without three dipole-dipole coupled electron donors. What's more, the dipole-dipole coupling strength \(J\) can greatly inhibit the RRRs with three dipole-dipole coupled electron donors, which indicates the quantum coherence generated by three coupled donors is an efficient approach to suppress RRR, and it is different from the quantum coherence mentioned by Marlan O. Scully [PRL 104, 207701 (2010)]. This presented scheme may propose some regulating strategies for efficient conversion efficiency via the suppressed RRR.
\begin{description}
\item[PACS numbers]42.50.Gy; 42.50.Ct; 32.80.Qk
\item[Keywords]Quantum photocell; radiative recombination rate (RRR); photon-to-charge efficiency
\end{description}
\end{abstract}
\maketitle
\section{Introduction}

The photon-to-charge conversion efficiency\cite{1} is an important aspect of photocell. However, the radiative upward transition and its reversal, the radiative downward transition coexist simultaneously, which produced the detailed balance limit\cite{2} in 1961 by Shockley and Queisser. And the radiative recombination has been considered as the fundamental limit\cite{2} on the conversion efficiency and is accepted widely in the artificial light-harvesting systems. But beyond that, other energy loss processes, such as surface reflection, internal resistance, thermalization losses, unabsorbed photons with energy less than band gap\cite{3} still exit in the photocell. However, they are regarded as unessential energy loss processes and many of them can be minimized by appropriately designed structures\cite{4,5,6,7,8}, such as multi-junction\cite{9,10,11} and intermediate band photocells \cite{12,13,14,15,16}, etc..

Recently, theoretical and experimental studies\cite{17,18,19} have demonstrated that the quantum coherence can alter the conditions of the detailed balance between the absorption and radiative recombination, and thereby the suppressed radiative recombination enhances the conversion efficiency in quantum photocells\cite{20,21,22,23}. One of the possible ways suggested by Scully\cite{17} is to cancel the emission processes via the quantum coherence induced by an external source\cite{18}. Consequently, the quantum coherence of the delocalized donor states alters the conditions for the thermodynamic detailed balance, and then brings out the enhanced efficiency in the photocell\cite{20,21,22,23}.

With the suppressed RRR in mind, this scheme focuses on the RRR in a proposed quantum photocell with three electron donors which can simulate the behavior of triple-junction photocells\cite{24,25}. The results indicate some positive significance and encouraging trends to photon-to-charge efficiency under the conditions with or without three dipole-dipole coupled electron donors.

The work is organized as follows, in section 2, we describe the quantum photocell model with three electron donors. And we present the results and the corresponding discussions regarding the RRR and possible experimental realization in section 3. A concise summary is given in the final section.

\section{Quantum photocell model with three electron donors}

\begin{figure}
\centerline{\includegraphics[width=0.45\columnwidth]{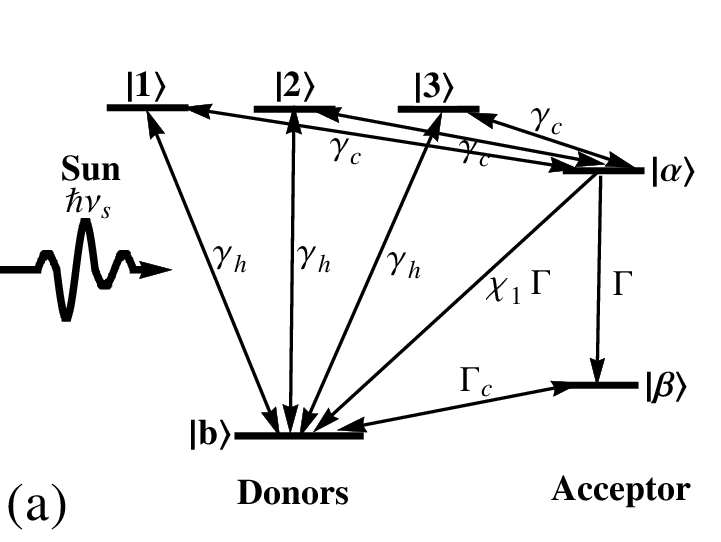}\includegraphics[width=0.45\columnwidth]{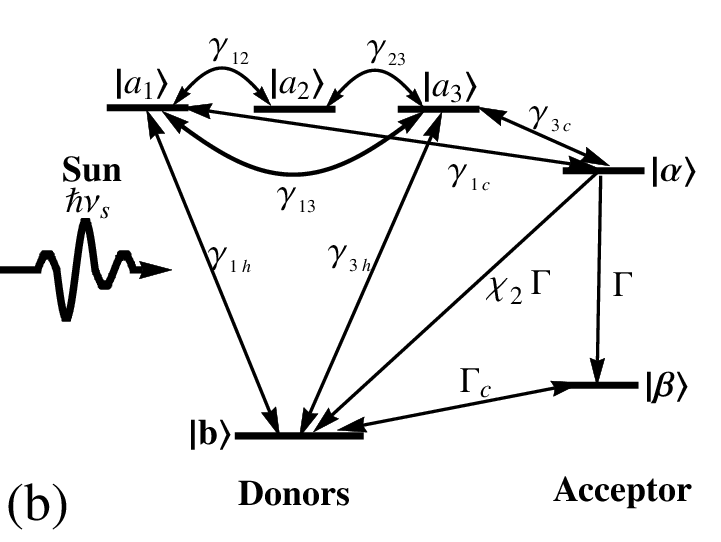}}
\caption{Schematic diagram: quantum photocell models with the acceptor and (a) three uncoupled donors, (b) three dipole-dipole coupled donors. Solar radiation drives electron transport between the valence band (VB) state \(|b\rangle\) and the conduction band (CB) state \(|i\rangle_{(i=1,2,3)}\) in Fig.(a). Transitions between levels \(|i\rangle_{(i=1,2,3)}\) \(\leftrightarrow\) \(|\alpha\rangle\), \(|\beta\rangle\) \(\leftrightarrow\) \(|b\rangle\) are driven by ambient thermal phonons. Levels \(|\alpha\rangle\) and \(|\beta\rangle\) are connected to a load. The three degenerate excited levels in Fig.1(a) split into Fig.1(b) because of the couplings between three donors, and the dark level \(|a_{2}\rangle\) is optically forbidden and has no electron transfer path to the acceptor \(|\alpha\rangle\).}
\label{f1}
\end{figure}

Proceeding with the analysis, we consider a quantum photocell model with the conduction band (CB) states \(|i\rangle_{(i=1,2,3)}\) and the valence band (VB) state \(|b\rangle\) [depicted in Fig.\ref{f1}(a)] as the donors. Level \(|\alpha\rangle\) and \(|\beta\rangle\) connecting to a load are assumed the acceptor molecule. The excitation of a molecule is simply modeled as a two-level system with the excited state \(|i\rangle_{(i=1,2,3)}\) and the ground state \(|b\rangle\). Then the excited electrons driven by solar radiation can be transferred to the acceptor molecule, the conduction reservoir state \(|\alpha\rangle\), with any excess energy radiated as a phonon into the ambient thermal phonons reservoirs. The excited electron is then assumed to be used to perform work, leaving the conduction reservoir state \(|\alpha\rangle\) decaying to the sub-stable state \(|\beta\rangle\) at a rate \(\Gamma\). The recombination between the acceptor and the donor is modeled as \(\chi_{i}\Gamma\) (i=1,2) in Fig.\ref{f1}, where \(\chi_{i} \) is the RRR, a dimensionless fraction. The recombination process brings the system back into the VB state \(|b\rangle\) without producing a work current, which could be a significant source of inefficiency. Finally, the state \(|\beta\rangle\) decays back to the VB state \(|b\rangle\) at a rate \(\Gamma_{c}\) and the cycle terminates. In Fig.\ref{f1}(a), the three donors are assumed to be identical and degenerate, and their three  uncoupled excited states \(|i\rangle_{(i=1,2,3)}\) have the same excitation levels \(E_{1}\)=\(E_{2}\)=\(E_{3}\)=E, and their transition dipole moments are aligned in the same direction, i.e., \(\vec{\mu}_{i}\)=e\(\langle i |\vec{r}|b\rangle_{(i=1,2,3)}\)=\(\vec{\mu}\), where \(\vec{r}=\vec{r}_{b}-\vec{r}_{i}\), and \(\vec{\mu}_{i}\) is located at \(\vec{r}_{i}\). The dipole-dipole interaction only exists in the nearest neighbors and the dipole-dipole couplings are denoted by J=\(\frac{1}{4\pi\epsilon\epsilon_{0}}[\frac{\vec{\mu}_{i}\cdot\vec{\mu}_{j}}{r^{3}}-\frac{3(\vec{\mu}_{i}\cdot \vec{r})(\vec{\mu}_{j}\vec{r})}{r^{5}}]\) between  \(|a_{1}\rangle \) and \(|a_{2}\rangle \), and \(|a_{2}\rangle \) and \(|a_{3}\rangle \) in Fig.\ref{f1}(b), but there is no coupling between \(|a_{1}\rangle \) and \(|a_{3}\rangle \). The strength of the dipole-dipole coupling J is much weaker than the excitation energy \(E- E_{b}=\hbar\omega\). The Hamiltonian of the three coupling donors can be written as

\begin{align}
\hat{H}=&\sum^{3}_{i=1} \hbar\omega {\hat{\sigma}}^{\dag}_{i} {\hat{\sigma}}_{i} + J(\hat{{\sigma}}^{-}_{1}{\hat{\sigma}}^{\dag}_{2}+{\sigma}^{-}_{2}{\sigma}^{\dag}_{3}+H.c.),
\end{align}

\noindent where H.c. means Hermitian conjugation, \( \hat{\sigma}^{\dag}_{i}\) and  \( \hat{\sigma}^{-}_{i}\) are the Pauli raising and lowering operators, respectively. The three single-excitation states of the above Hamiltonian are \(|a_{1}\rangle\) =\(\frac{1}{2}\)(\(|1\rangle\)+\(\sqrt{2}|2\rangle\)+\(|3\rangle\)), \(|a_{2}\rangle\) =\(\frac{1}{\sqrt{2}}\)(\(|1\rangle\)-\(|3\rangle\))
,  \(|a_{3}\rangle\) =\(\frac{1}{2}\)(\(|1\rangle\)-\(\sqrt{2}|2\rangle\)+\(|3\rangle\)), and their eigenvalues are obtained as
\(E_{a_{1}}\)=\(E\)+\(\sqrt{2}\)J, \(E_{a_{2}}\)=\(E\), \(E_{a_{3}}\)=\(E\)-\(\sqrt{2}\)J. The dynamics behaviors of the donors-acceptor systems can be described via the master equations in Eq.(\ref{eq.2}) and Eq.(\ref{eq.3}) without and with three dipole-dipole coupled donors as follows, respectively.

\begin{align}
\dot{\rho_{11}}={}& \gamma_{h}[n_{h}\rho_{bb}\!-\!(1+n_{h})\rho_{11}]\!+\!\gamma_{c}[n_{c}\rho_{\alpha\alpha}\!-\!(1+n_{c})\rho_{11}],\nonumber\\
\dot{\rho_{22}}={}& \gamma_{h}[n_{h}\rho_{bb}\!-\!(1+n_{h})\rho_{22}]\!+\!\gamma_{c}[n_{c}\rho_{\alpha\alpha}\!-\!(1+n_{c})\rho_{22}], \nonumber\\
\dot{\rho_{33}}={}& \gamma_{h}[n_{h}\rho_{bb}\!-\!(1+n_{h})\rho_{33}]\!+\!\gamma_{c}[n_{c}\rho_{\alpha\alpha}\!-\!(1+n_{c})\rho_{33}], ~~~~~\label{eq.2} \\
\dot{\rho_{\alpha\alpha}}={}&\gamma_{c}(1+n_{c})(\rho_{11}\!+\!\rho_{22}+\rho_{33})\!-\!3\gamma_{c}n_{c}\rho_{\alpha\alpha}\!-\!\Gamma(1+\chi_{1})\rho_{\alpha\alpha},\nonumber\\
\dot{\rho_{\beta\beta}}={}& \Gamma_{c}[N_{c}\rho_{bb}\!-\!(1+N_{c})\rho_{\beta\beta}]+\Gamma\rho_{\alpha\alpha},\nonumber
\end{align}

\noindent and

\begin{eqnarray}
&\dot{\rho}_{a_{1}a_{1}}=&\gamma_{1h}[n_{1h}\rho_{bb}-(1+n_{1h})\rho_{a_{1}a_{1}}]+\gamma_{12}[n_{12}\rho_{a_{2}a_{2}}
                        \nonumber\\&&-(1+n_{12}) \rho_{a_{1}a_{1}}+\gamma_{13}[n_{13}\rho_{a_{3}a_{3}}-(1+n_{13})\rho_{a_{1}a_{1}}] \nonumber\\&&+\gamma_{1c}[n_{1c}\rho_{\alpha\alpha}-(1+n_{1c})\rho_{a_{1}a_{1}}],\nonumber\\
&\dot{\rho}_{a_{2}a_{2}}=&\gamma_{12}[(1+n_{12})\rho_{a_{1}a_{1}}-n_{12}\rho_{a_{2}a_{2}}]+\gamma_{23}[n_{23}\rho_{a_{3}a_{3}} \nonumber\\
                        &&-(1+n_{23})\rho_{a_{2}a_{2}}],~~~~~~~~~~~~~~~~~~~~~~~~~~~~~~~~~~~~ \label{eq.3}\\
&\dot{\rho}_{a_{3}a_{3}}=&\gamma_{3h}[n_{3h}\rho_{bb}-(1+n_{3h})\rho_{a_{3}a_{3}}]+\gamma_{23}[(1+n_{23})\rho_{a_{2}a_{2}}
                        \nonumber\\&&-n_{23}\rho_{a_{3}a_{3}}]+\gamma_{13}[(1+n_{13})\rho_{a_{1}a_{1}}-n_{13}\rho_{a_{3}a_{3}}] \nonumber\\&&+\gamma_{3c}[n_{3c}\rho_{\alpha\alpha}-(1+n_{3c})\rho_{a_{3}a_{3}}],\nonumber\\
&\dot{\rho}_{\alpha\alpha}=&\gamma_{1c}[(1+n_{1c})\rho_{a_{1}a_{1}}-n_{1c}\rho_{\alpha\alpha}]+\gamma_{3c}[(1+n_{3c})\rho_{a_{3}a_{3}}
                        \nonumber\\&&-n_{3c}\rho_{\alpha\alpha}]-\Gamma(1+\chi_{2})\rho_{\alpha\alpha},\nonumber\\
&\dot{\rho}_{\beta\beta}=&\Gamma\rho_{\alpha\alpha}+\Gamma_{c}[\emph{N}_{c}\rho_{bb}-(1+\emph{N}_{c})\rho_{\beta\beta}],\nonumber
\end{eqnarray}

\noindent where \(n_{h}\)=\(\frac{1}{exp(\frac{E-E_{b}}{K_{B}T_{s}})-1}\), \(n_{ih(i=1,3)}\)=\(\frac{1}{exp(\frac{E_{a_{i}}-E_{b}}{K_{B}T_{s}})-1}\) describe the average numbers of photon with frequencies matching the transition energies from the VB state \(|b\rangle\) to the CB states \(|i\rangle_{(i=1,2,3)}\) in Fig.\ref{f1}(a), and  \(|a_{i}\rangle_{(i=1,3)}\) in Fig.\ref{f1}(b) at the temperature \(T_{s}\)=(300 +\(\Delta\))K, where \(\Delta\) stands for the temperature difference. \(n_{c}\)=\(\frac{1}{exp(\frac{E-E_{\alpha}}{K_{B}T_{s}})-1}\) and  \(n_{ic(i=1,3)}\)=\(\frac{1}{exp(\frac{E_{a_{i}}-E_{\alpha}}{K_{B}T_{s}})-1}\) are the thermal occupation numbers of ambient phonons at temperature \(T_{s}\).  \(N_{c}\)=\(\frac{1}{exp(\frac{E_{\beta}-E_{b}}{k_{B}T_{s}})-1}\) is the corresponding thermal occupation number at the ambient temperature \(T_{s}\) with energy gap (\(E_{\beta}\)-\(E_{b}\)). \(n_{12}\), \(n_{13}\), and \(n_{23}\) represent the corresponding thermal occupations at the ambient temperature \(T_{s}\) with energy gaps (\(E_{a_{1}}\)-\(E_{a_{2}}\)), (\(E_{a_{1}}\)-\(E_{a_{3}}\)), and (\(E_{a_{2}}\)-\(E_{a_{3}}\)), respectively. The rates in Eq.(\ref{eq.2}) and Eq.(\ref{eq.3}) lead to a Boltzmann distribution for the level population \(|\alpha\rangle\) (\(p_{\alpha}\)=\(exp(-\frac{E_{\alpha}-\mu_{\alpha}}{k_{B}T_{s}})\)), \(\mu_{\alpha}\) is defined as the chemical potential of lead \(\alpha\)) when the thermal averages for the photon and phonon reservoirs are in a common temperature. We consider the initial condition to be a fully occupied ground state\cite{22}, i.e., \(\rho_{bb}\) = 1.

\section{Summary and Discussion}

In what follows, we calculate the steady solutions of Eq.(\ref{eq.2}) and Eq.(\ref{eq.3}) for the RRRs, \(\chi_{1}\) and \(\chi_{2}\).  Considering the cumbersome expressions for \(\chi_{1}\) and \(\chi_{2}\), we follow the numerical approach to carry out the discussion. And we use the following parameters\cite{21,22}, \(E_{\alpha}-\mu_{\alpha}\)=0.10 ev, \(E-E_{\alpha}\)=\(E_{\beta}-E_{b}\)=0.20 ev, \(\gamma_{c}\)=6 Mev, \(\Gamma\)=0.40 ev, \(\Gamma_{c}\)=0.15 ev. And other parameters are \(\gamma_{h}\)=0.62 ev in Fig.\ref{f1}(a), and \(\gamma_{1h}\)=0.62 ev ,\(\gamma_{3h}\)=0.45 ev, \(\gamma_{1c}\)=\(\gamma_{3c}\) =0.15\(\ast\)\((\frac{3}{2}\) +\(\sqrt{2}\)) ev, J=0.10 ev, \(\gamma_{12}\)=\(\gamma_{23}\) =\(\frac{1}{2}\)\(\gamma_{13}\)=0.15\(\ast\)\(\sqrt{2}\) ev in Fig.\ref{f1}(b).

\begin{figure}
\centerline{\includegraphics[width=0.45\columnwidth]{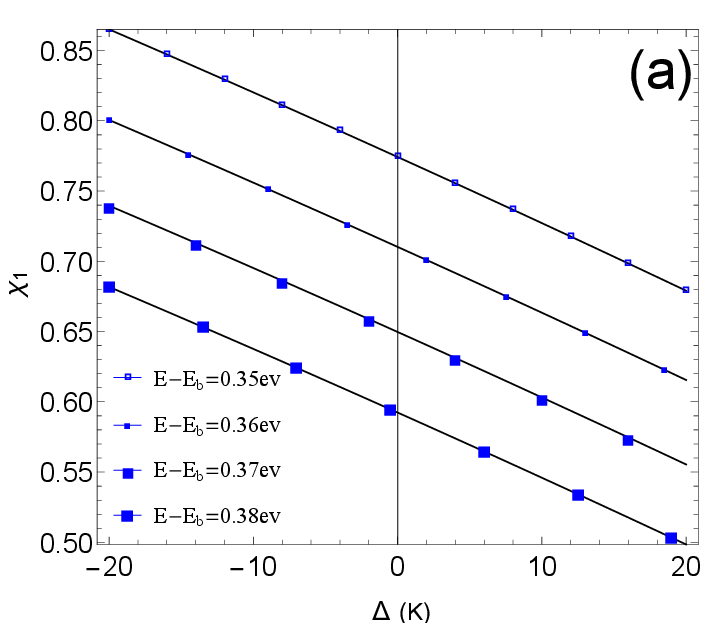}\includegraphics[width=0.46\columnwidth]{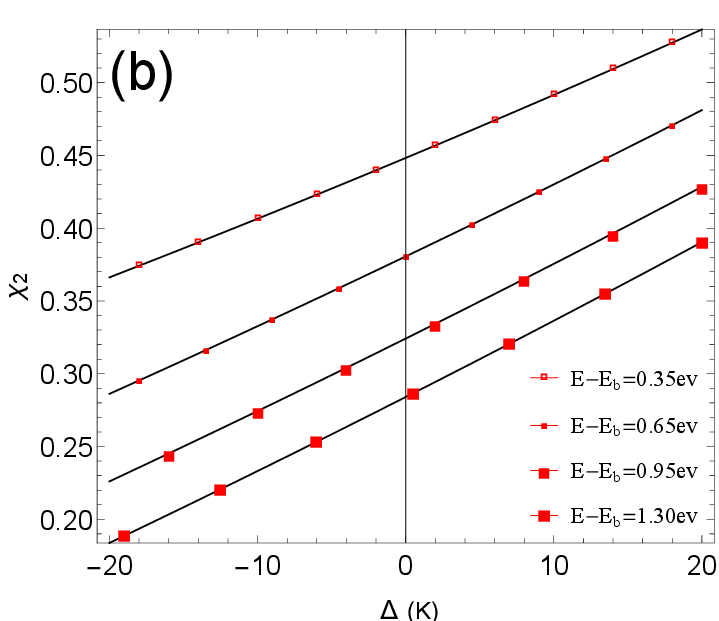}}
\caption{(Color online) (a) RRR \(\chi_{1}\) without three coupled donors, (b) RRR \(\chi_{2}\) with three coupled donors as a function of the temperature difference \(\Delta\)(K) with different gaps (\(E-E_{b}\)). }
\label{f2}
\end{figure}

Fig.\ref{f2} plots the RRR (\(\chi_{i}\)) as a function of the temperature difference \(\Delta\)(K) with the control-parameters being energy gaps (\(E-E_{b}\)). Fig.\ref{f2} (a) shows a linear decreasing feature dependent the temperature difference \(\Delta\)(K) and energy gaps,
(\(E-E_{b}\)), and RRR (\(\chi_{1}\)) declines from 0.77 to 0.59 with the damping amplitude about 0.07 at room temperature (\(\Delta\)=0). These results indicate that the increasing ambient temperature and energy gaps (\(E-E_{b}\)) can block the radiative recombination from the acceptor to donors without three coupled electron donors. Therefore, the transported electrons to VB via the radiative recombination become less and more excited electrons are transported to perform useful work. These features suggest that the quantum photocell with a wider band gap but without three coupled electron donors operates at a slight higher temperature will bring out more excited electrons and achieve high conversion efficiency. However, we should also take account of the fact that, increasing the band gap and ambient temperature will cause less absorption and much more phonon-electron scattering. Therefore, Fig.\ref{f1}(a) suggests that the room temperature and a proper energy gap (\(E-E_{b}\)) are the optimal operating conditions for the photocell without three coupled electron donors in the real operating environment.

However, a contrary result appears in Fig.\ref{f2}(b) with three dipole-dipole coupled donors. RRRs (\(\chi_{2}\)) show the linear increasing character dependent the temperature difference \(\Delta\)(K), but they still decrease with the increments of the energy gaps (\(E-E_{b}\)).
What's more, the values of RRRs in Fig.\ref{f2}(b) are much smaller than those in Fig.\ref{f2}(a) at room temperature (\(\Delta\)=0). The reason comes from their difference, i.e., the transport electrons transiting to the VB \(|b\rangle\) are inhibited in Fig.\ref{f2}(b), due to the quantum coherence generated by their dipole-dipole interactions which is different from that mentioned by Marlan O. Scully\cite{18}. But the higher ambient temperature can weaken the quantum coherence and incurs the increment of RRRs \(\chi_{2}\) in Fig.\ref{f2}(b). Hence, just taking three dipole-dipole coupled donors into account, the quantum coherence generated by three dipole-dipole coupled donors can enhance the photon-to-charge efficiency via the suppressed RRRs \(\chi_{2}\) at a proper circumstance temperature. Although, Fig.\ref{f2}(b) indicates that a slightly large gap (\(E-E_{b}\)) in the p-n junction can diminish the possibility of the electronic radiative recombination, less absorption solar photons due to a wider band gap is an undesirable fact in the photocell. So, a proper gap (\(E-E_{b}\)) benefits to the suppressed RRR and profits the conversion efficiency in the actual photocell design.

\begin{figure}
\centerline{\includegraphics[width=0.44\columnwidth]{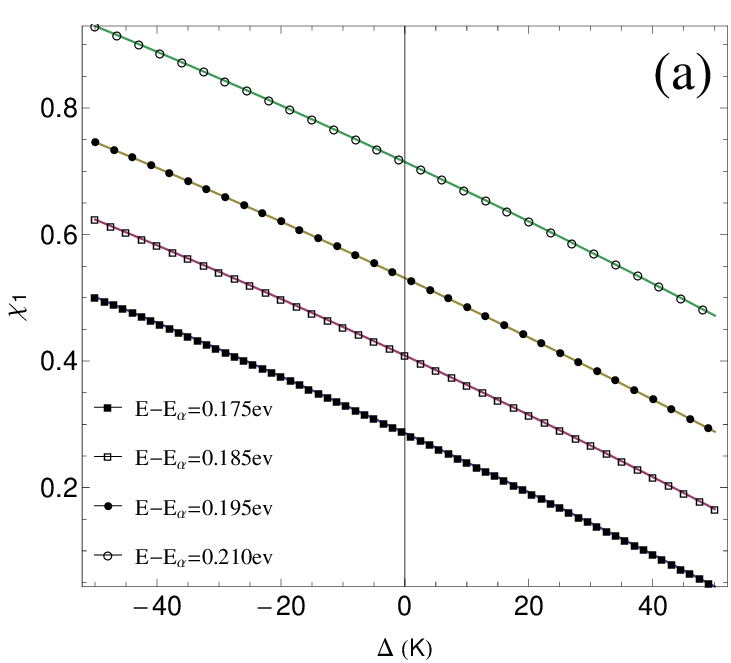}\includegraphics[width=0.45\columnwidth]{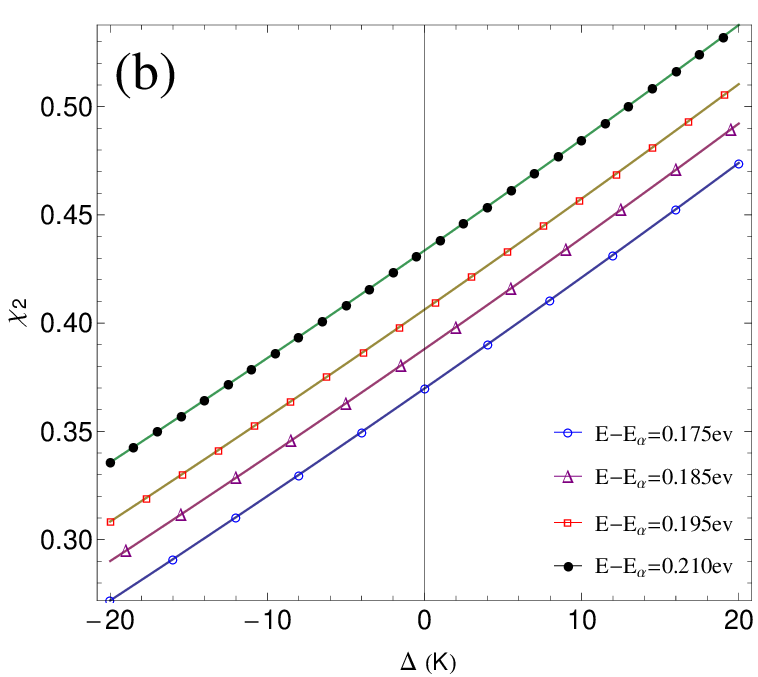}}
\caption{(Color online) (a) RRR \(\chi_{1}\) without three coupled donors, (b) RRR \(\chi_{2}\) with three coupled donors as a function of the temperature difference \(\Delta\)(K)  with different gaps (\(E-E_{\alpha}\)). }
\label{f3}
\end{figure}

In photocell, the gaps (\(E_{i,(i=1,2,3)}-E_{\alpha}\)) between the donors and acceptor molecular, energy gap (\(E-E_{\alpha}\)) may be encountered during the investigation to photoelectric conversion pro cess. In the following, the curves show its influence on the RRRs with the gaps (\(E-E_{b}\))=0.38 ev, and 0.7 ev in Fig.\ref{f3}(a) and (b), respectively, and other parameters are the same to those in Fig.\ref{f2}.

The curves show the similar features dependent ambient temperatures, i,e., the linear decrement RRR \(\chi_{1}\) but linear increment RRR \(\chi_{2}\) dependent ambient temperatures are displayed in Fig.\ref{f3} (a) and (b), respectively. Not only that, but the RRRs \(\chi_{1}\) and \(\chi_{2}\) are both enhanced by the increasing gaps (\(E-E_{\alpha}\)) in Fig.\ref{f3} (a) and (b). The results indicate that more electrons are excited under a higher ambient temperature without three dipole-dipole coupling donors shown in Fig.\ref{f3} (a), but the quantum coherence is destroyed by the increasing ambient temperature, which incurs the increasing \(\chi_{2}\) in Fig.\ref{f3}(b). And it is hard for the excited electrons to transport to the external load with a larger gap (\(E-E_{\alpha}\)), but radiative to the VB state \(|b\rangle\) ultimately. So, as shown in Fig.\ref{f3}, the increasing (\(E-E_{\alpha}\)) enhances RRR both in Fig.\ref{f3} (a) and (b). Comparing their amplitude values at room temperature in Fig.\ref{f2} and Fig.\ref{f3}, \(\chi_{2}\) is more severely suppressed than \(\chi_{1}\), which demonstrates the physical facts that the inhibited RRRs can be easily achieved with three dipole-dipole coupled donors than those without three coupled donors. The results manifest that the quantum coherence can effectively suppress the radiative radiation to the VB state in this quantum photocell with three coupled donors.

\begin{figure}
\centerline{\includegraphics[width=0.45\columnwidth]{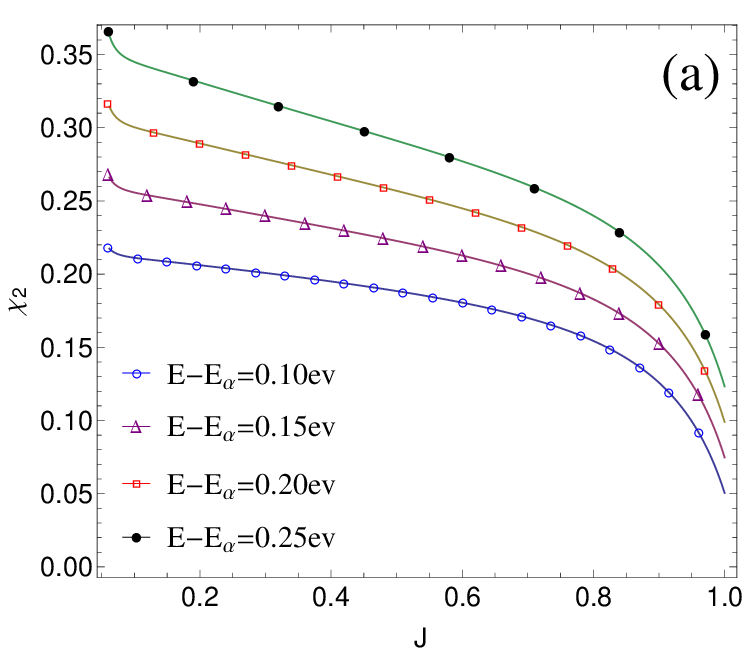}\includegraphics[width=0.44\columnwidth]{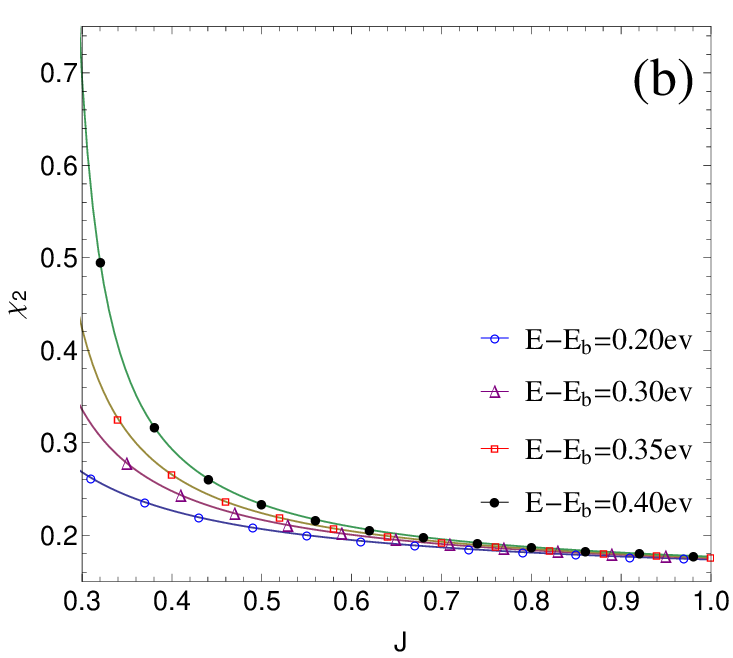}}
\caption{(Color online) The RRR \(\chi_{2}\) with three coupled donors as a function of the electrostatic dipole-dipole coupling strength \(J\) with different energy gaps at room temperature.(a)(\(E-E_{b}\))=1.6 ev, (b) (\(E-E_{\alpha}\))=0.05 ev, and other parameters take the same values to those in Fig.2. }
\label{f4}
\end{figure}

As mentioned before, RRR dependent the ambient temperature and gaps are discussed in this quantum photocell with or without dipole-dipole interaction in Fig.\ref{f2} and Fig.\ref{f3}. And RRRs in this photocell with dipole-dipole interaction are always less than those without dipole-dipole interaction. In fact, the increasing electrostatic dipole-dipole coupled strength \(J\) can bring out the stepped-up quantum interference between the different donors. Therefore, the quantum interference between three dipole-dipole coupled donors can be described by the electrostatic dipole-dipole coupled strength \(J\). Hence, the electrostatic dipole-dipole coupling strength \(J\) should be paid some attention in the photocell with three dipole-dipole coupled interaction.

Fig.\ref{f4} shows RRRs \(\chi_{2}\) versus the electrostatic dipole-dipole coupling strength \(J\) at room temperature with gaps (\(E-E_{\alpha}\))and (\(E-E_{b}\)) being the controlled parameters in Fig.\ref{f4}(a) and (b), respectively. It shows that \(\chi_{2}\) monotonically decreases with the increasing \(J\), and the less gaps (\(E-E_{\alpha}\)), (\(E-E_{b}\)) generate the less RRR \(\chi_{2}\) both in Fig.\ref{f4} (a) and (b). However, the sharp decrement \(\chi_{2}\) appears about in the range 0.8\(<\) \(J\) \( <\)1 in Fig.\ref{f4}(a) while about in the range 0.3\(<\) \(J\) \( <\)0.5 in Fig.\ref{f4}(b), and the final values of \(\chi_{2}\) are in close proximity to 0.05 in Fig.\ref{f4}(a), while infinitesimally approach to 0.2 in Fig.\ref{f4}(b). These results indicate that the increasing quantum interference between three dipole-dipole coupled donors and less gaps (\(E-E_{\alpha}\)) and (\(E-E_{b}\)) can inhibit the RRR \(\chi_{2}\) ultimately, and the RRR can be suppressed to a minimum but can't be canceled out. The physical significance is that the coexisting of the radiative upward and downward transition are still in the quantum photocell\cite{17} even if it can be regulated by the quantum coherence.

Up to now, we have investigated the features of RRR dependent the ambient temperature, gaps and the electrostatic coupling strength \(J\) in this quantum photocell with three electron donors. Before concluding this paper, we would like to point out some items. First of all, the quantum photocell with three electron donors is just a precedent for multi-donors, and has the similar physical regime to those with multi-donors. Secondly, two type of energy gaps discussed here display some significant results with respect to the RRR. The results generated by the gaps within the donors indicate that semiconductor materials with appropriate energy gaps can effectively inhibit radiative recombination. Therefore, seeking a semiconductor material with suitable band gap not only to absorb more solar photons but also to suppress RRR, which deserves further experimental investigation. And the gaps between the donors and acceptor may be another experimental domain according our results, and best-effort to reduce this gap is a possible way to the efficient photovoltaic conversion. The last point is how to align the donors of photocells for a robust electrostatic dipole-dipole coupling strength \(J\) in the manufacturing process, which introduces a different quantum coherence to suppress RRR in this scheme. The scenario proposed here may be a different approach for efficient photon-to-charge conversion and deserve further experimental investigation.

\section{Conclusion}

To summarize, in this work we explored the RRR dependent ambient temperature, gaps and the electrostatic coupling strength of the dipole-dipole coupling \(J\) in a quantum photocell system with three electron donors. It showed that the RRRs can be suppressed by the increasing ambient temperatures and gaps within the donors, but enhanced by gaps between the donors and acceptor without three dipole-dipole coupled electron donors. However, the increments of ambient temperatures and gaps between the donors and acceptor promote the RRRs, while the gaps within the donors and electrostatic dipole-dipole coupling strength \(J\) can greatly suppress the RRRs with three dipole-dipole coupled electron donors. Apart from the gap and ambient temperature regulation, this scheme propose a different quantum coherence to suppress the RRR in this quantum photocell, and these results may attract some experimental interesting.

\section*{Conflict of Interest}
The authors declare that they have no conflict of interest. This article does not contain any studies with human participants or animals performed by any of the authors. Informed consent was obtained from all individual participants included in the study.
\begin{acknowledgments}
We thank the financial supports from the National Natural Science Foundation of China ( Grant Nos. 61205205 and 61565008 ), and
the General Program of Yunnan Applied Basic Research Project, China ( Grant No. 2016FB009 ).
\end{acknowledgments}

\section*{Author contribution statement}
S. C. Zhao conceived the idea. J. Y. Chen performed the numerical computations and wrote the draft, and S. C. Zhao did the analysis and revised the paper.




\bibliographystyle{0}

\begin{thebibliography}{00}
\bibitem{1} P. W$\ddot{u}$rfel, {\it Physics of Solar Cells}, (Wiley-VCH, Berlin), (2009).
\bibitem{2} W. Shockley and H. J. Queisser, {\it J. Appl. Phys.} {\bf 32}, 510 (1961).
\bibitem{3} M. A. Green, K. Emery, Y. Hishikawa, W. Warta, E. D. Dunlop, {\it Solar cell efficiency tables (version 48).Prog. Photovolt: Res. Appl.} {\bf 24}, 905, (2016).
\bibitem{4} U. W$\ddot{u}$rfel, M. Thorwart and E. R. Weber, {\it Quantum Efficiency in Complex Systems, Part II: From Molecular Aggregates to
            Organic Solar Cells, in Semiconductors and Semimetals},(vol. 85, Academic Press, San Diego), (2011).
\bibitem{5} O. D. Miller, E. Yablonovitch and S. R. Kurtz, {\it IEEE J. Photovoltaics}, {\bf 2 }, 303 (2012).
\bibitem{6} F. H. Alharbi, {\it J. Phys. D: Appl. Phys.}, {\bf 46}, 125102 (2013).
\bibitem{7} F. H. Alharbi and S. Kais, {\it Renewable Sustainable Energy Rev.}, {\bf 43}, 1073 (2015).
\bibitem{8} M. Daryani, A. Rostami, G. Darvish, M. K. Morravej Farshi, {\it Opt. Quant. Electron.}, {\bf 49}, 255 (2017).
\bibitem{9} C. H. Henry, {\it J. Appl. Phys.}, {\bf 51 }, 4494 (1980).
\bibitem{10} A. Luque, P. G. Linares, E. Antolin, E. Canovas, C. D. Farmer, C.R. Stanley and A. Marti, {\it  Appl. Phys. Lett.}, {\bf 96}, 013501 (2010).
\bibitem{11} T. Nozawa and Y. Arakawa, {\it Appl. Phys. Lett.}, {\bf 98}, 171108 (2011).
\bibitem{12} A. Luque and A. Marti, {\it Phys. Rev. Lett.}, {\bf 78}, 5014 (1997).
\bibitem{13} J. Li, M. Chong, J. Zhu, Y. Li, J. Xu, P. Wang, Z. Shang, Z. Yang, R. Zhu and X. Cao, {\it Appl. Phys. Lett.}, {\bf 60}, 2240 (1992).
\bibitem{14} J. Bruns, W. Seifert, P. Wawer, H. Winnicke, D. Braunig and H. G. Wagemann, {\it Appl. Phys. Lett.}, {\bf 64}, 2700 (1994).
\bibitem{15} H. Kasai, H. Matsumura, {\it Sol. Energy Mater. Sol. Cells}, {\bf 48}, 93 (1997).
\bibitem{16} A. Luque, {\it J. Appl. Phys.}, {\bf 11}, 031301 (2011).
\bibitem{17} K. E. Dorfman, M. B. Kim, A. A. Svidzinsky, {\it Phys. Rev. A},{\bf 84}, 053829 (2011)
\bibitem{18} M. O. Scully, {\it Phys. Rev. Lett.}, {\bf 104}, 207701 (2010).
\bibitem{19} A. A. Svidzinsky, K. E. Dorfman and M. O. Scully, {\it Phys. Rev. A}, {\bf 84}, 053818 (2011).
\bibitem{20} M. O. Scully, K. R. Chapin, K. E. Dorfman, M. B. Kim and A. Svidzinsky, {\it Proc. Natl. Acad. Sci.}, {\bf 108}, 15097 (2011).
\bibitem{21} K. E. Dorfman, D. V. Voronine, S. Mukamel and M. O. Scully, {\it Proc. Natl. Acad. Sci.}, {\bf 110}, 2746 (2013).
\bibitem{22} C. Creatore, M. A. Parker, S. Emmott and A. W. Chin, {\it Phys. Rev. Lett.}, {\bf 111}, 253601 (2013).
\bibitem{23} K. E. Dorfman, K. E. Dorfman and M. O. Scully, {\it Coherent Optic Phenom},{\bf 1}, 42 (2013).
\bibitem{24} R. R. King, D. C. Law, K. M. Edmondson, C. M. Fetzer, et al., {\it Appl. Phys. Lett.}, {\bf 90}, 183516 (2007).
\bibitem{25} R. R. King, D. C. Law, K. M. Edmondson, et al., {\it Advances in Opto-Electronics},{\bf 2007}, 29523 (2007).
\end{thebibliography}

\end{document}